\documentclass[preprint,preprintnumbers, prd, floatfix, superscriptaddress,nofootinbib]{revtex4-1}
\usepackage{epsfig}
\usepackage{subfigure}
\usepackage{dcolumn}
\usepackage{bm}
\usepackage[usenames ,dvipsnames]{xcolor}
\usepackage{slashed}
\usepackage{graphicx,color}
\begin{document}
\title{Testing the light scalar meson as a non-$q\bar q$ state\\ 
in semileptonic $D$ decays}

\author{Yu-Kuo Hsiao}
\email{Corresponding author: yukuohsiao@gmail.com}
\affiliation{School of Physics and Information Engineering, 
Shanxi Normal University, Taiyuan 030031, China}

\author{Shu-Qi Yang}
\email{yshuqi0313@163.com}
\affiliation{School of Physics and Information Engineering, 
Shanxi Normal University, Taiyuan 030031, China}

\author{Wen-Juan Wei}
\email{snza12321@163.com}
\affiliation{School of Physics and Information Engineering, 
Shanxi Normal University, Taiyuan 030031, China}

\author{Bai-Cian Ke}
\email{baiciank@ihep.ac.cn}
\affiliation{School of Physics and Microelectronics,
Zhengzhou University, Zhengzhou, Henan 450001, China}

\date{\today}

\begin{abstract}
While the light scalar mesons ($S_0$) are considered to be 
either ordinary $q\bar q$ or exotic tetraquark states,  
we investigate the semileptonic decays $D\to S_0 e^+\nu_e$, by taking into account
the resonant effects of $S_0\to M_1 M_2$, 
where $S_0=a_0(980)$, $f_0(980)$, and $f_0(500)/\sigma_0$, and
$M_{1(2)}$ represents a pseudoscalar meson.
For the first time, the $D\to S_0$ form factors in the two quark structures are both presented.
Subsequently, we calculate
${\cal B}(D_s^+\to \sigma_0 e^+\nu_e,\sigma_0\to\pi^+\pi^-)=(20.3\pm 1.8\pm 0.5)\times 10^{-4}$
in the $q\bar q$ structure, showing significant $9\sigma$ deviations
from the experimental upper limit of $3.3\times 10^{-4}$. In contrast,
${\cal B}(D_s^+\to \sigma_0 e^+\nu_e,\sigma_0\to\pi^+\pi^-)=(0.58^{+1.43}_{-0.57}\pm 0.01)\times 10^{-4}$
in the $q^2\bar q^2$ structure is within the allowed experimental range.
Clearly, the light scalar meson is tested as a non-$q\bar q$ state.
We hence demonstrate a highly sensitive new approach 
for exploring the true nature of scalar mesons, 
which can be extended to non-leptonic $D$ decays as well as $B$ meson decays.
\end{abstract}

\maketitle
\section{introduction}
Exotic particles have been recently discovered
but are confined to be charmful cases~\cite{Belle:2003nnu,Aaij:2020hon,LHCb:2021vvq,LHCb:2015yax}.
Their exotic nature can be interpreted either 
as compact multiquark bound states due to gluon exchanges or 
as loosely bound hadron-hadron molecules 
due to residual nuclear force~\cite{Chen:2022asf,Pelaez:2015qba,Guo:2017jvc}. 
Further clarification can be challenging, 
requiring an improved understanding of quark confinement 
in quantum chromodynamics (QCD) and the low-energy strong interaction.

Even though charmful exotic states have been richly observed,  
such as $X(3872)\sim c\bar c u\bar u(d\bar d)$~\cite{Belle:2003nnu}, 
$X_{0,1}(2900)^0\sim \bar c d \bar s u$~\cite{Aaij:2020hon}, 
$T_{cc}(3875)^+\sim cc u\bar d$~\cite{LHCb:2021vvq}, and
${\cal P}_c(4380,4450)^+$ $\sim c\bar c uud$~\cite{LHCb:2015yax},
the existence of their charmless counterparts has not been confirmed yet. 
An example highlighting this is the resonant decay: $B^+\to \bar p \Theta(1710)^{++}$,
$\Theta(1710)^{++}\to pK^+$ with $\Theta(1710)^{++}$ composed of $uud\bar s u$,
measured with an upper limit of the branching fraction: ${\cal B}<9.1\times 10^{-8}$~\cite{pdg}. 
This raises a serious question: can charmless exotic states,
as proposed by the quark model~\cite{Gell-Mann:1964ewy,zweig}, be discovered?

Light scalar meson ($S_0$), identified as $f_0/f_0(980)$, $a_0/a_0(980)$, 
$\sigma_0/f_0(500)$, or $\kappa/K^*_0(700)$, has long been considered 
a compact four-quark $q^2\bar q^2$ bound state~\cite{Jaffe:1976ig,Jaffe:1976ih,
Close:2002zu,Pelaez:2003dy,Maiani:2004uc,Amsler:2004ps,Jaffe:2004ph,
Achasov:2005hm,tHooft:2008rus,Fariborz:2009cq,Weinberg:2013cfa,Agaev:2018fvz}
or a loosely-bound meson-pair molecule~\cite{Weinstein:1990gu, Branz:2007xp, Baru:2003qq,
Dai:2012kf, Dai:2014lza, Sekihara:2014qxa, Yao:2020bxx, Wang:2022vga}.
Either of these structures consists of $q\bar q q\bar q$, making
$S_0$ a promising charmless exotic tetraquark candidate. Alternatively, 
some proposals suggest it to be an ordinary p-wave $q\bar q$ meson~\cite{Anisovich:2000wb,
Bediaga:2003zh,Aliev:2007uu,Colangelo:2010bg,Shi:2015kha,Soni:2020sgn,Klempt:2021nuf}.
Without excluding this possibility, the existence of a charmless exotic tetraquark 
cannot be definitely confirmed. 
On the other hand, it has never been doubted that 
$X_{0,1}(2900)^0$ and $T_{cc}(3875)^+$ are exotic states,
based on the fact that the experiments have clearly ruled them out 
as ordinary mesons~\cite{Aaij:2020hon,LHCb:2021vvq}.

Utilizing the scattering amplitude in meson-meson interactions~\cite{Oller:1997ng},  
the compositeness relation $X$ is introduced to assess 
the molecule structure of the light scalar meson~\cite{Weinberg:1962hj,
Dai:2012kf,Guo:2015daa,Sekihara:2014qxa,Sekihara:2014qxa,Wang:2022vga}.
The elementariness relation $Z=1-X$ is also introduced, 
estimating the ordinary $q\bar q$ and compact $q^2\bar q^2$ components, 
but barely distinguishing between them. 
Thus, the $q\bar q$ structure remains unexcludible. 

The weak decays of $B$ mesons and $D$ mesons can serve as an accurate probe 
for the $Z$-included $q\bar q$ and $q^2\bar q^2$ structures 
of the light scalar meson~\cite{Wang:2009azc,Stone:2013eaa,
Achasov:2021dvt,Achasov:2020aun,Cheng:2017fkw,Achasov:2020qfx,Cheng:2002ai},
thereby distinguishing between them.
Particularly, semileptonic decays $D\to S_0 e^+\nu_e$ followed by $S_0\to M_1 M_2$
offer a favorable platform. 
Accordingly, BESIII has recently measured the semileptonic $D_s^+$ decays~\cite{BESIII:2023wgr,BESIII:2021drk}.
The $D_s^+$ to $f_0$ transition form factor has also been determined as
$f^+(0)=0.518\pm 0.018\pm 0.036$~\cite{BESIII:2023wgr}.
However, it is poorly understood how these measurements can be used to explore 
the true nature of the light scalar mesons.

The mixing of light scalar mesons and the associated mixing angle 
facilitate a connection of all form factors in $D\to S_0 e^+\nu_e, S_0\to M_1 M_2$ decays
under the $SU(3)$ flavor  $[SU(3)_f]$ symmetry.
This introduces another intriguing aspect to enhance our exploration.
Considering that the mixing angle $|\theta|$ is commonly estimated 
to be around $20^\circ$ for the $q\bar{q}$ structure~\cite{Klempt:2021nuf}
and less than $10^\circ$ for the $q^2\bar q^2$ structure~\cite{Maiani:2004uc},
the form factors corresponding to these two distinct quark contents 
could result in significant differences in the branching fractions.

In this report, we aim to establish the connection among the form factors 
in the semileptonic decays of $D_s^+$ and $D^{0(+)}$. 
We will re-extract $f^+(0)$ in the $D_s^+\to f_0 e^+\nu_e, f_0\to\pi^+\pi^-$ decay, 
ensuring its alignment with the experimental determination. 
Through the computation of branching fractions associated with 
the two distinct quark structures and a subsequent comparison with experimental data, 
we anticipate demonstrating a strong preference for one quark content over the other.

\section{Formalism}
According to the effective Hamiltonians of
the quark-level $c\to q e^+ \nu_e$ decays~\cite{Buras:1998raa},
the resonant amplitude of $D\to S_0 e^+ \nu_e,S_0\to M_1 M_2$ 
can be presented as
\begin{eqnarray}\label{amp1}
{\cal M}(D\to S_0 e^+ \nu_e,S_0\to M_1 M_2)&=&\frac{G_F}{\sqrt 2}V_{cq}
\langle M_1 M_2|\bar q\gamma_\mu(1-\gamma_5) c|D\rangle_{\rm res}\bar u_\nu\gamma^\mu(1-\gamma_5) v_e\,,
\end{eqnarray}
where $G_F$ is the Fermi constant, $V_{cq}$ 
the Cabibbo-Kobayashi-Maskawa (CKM) matrix element, and $q$ denotes $s$ or $d$. 
The subscript ``res'' denotes the resonant $D\to S_0\to M_1 M_2$ transition,
whose matrix elements can be expressed as
\begin{eqnarray}\label{Amp2}
\langle M_1 M_2|\bar q\gamma_\mu(1-\gamma_5) c|D\rangle_{\rm res}=
\langle M_1 M_2|S_0\rangle D_{S_0}^{-1}\langle S_0|\bar q\gamma_\mu(1-\gamma_5)c|D\rangle\,.
\end{eqnarray}
In Eq.~(\ref{Amp2}), 
$\langle S_0|\bar q\gamma_\mu(1-\gamma_5) c|D\rangle$ represents the $D\to S_0$ transition,
$D_{S_0}^{-1}$ accounts for $S_0$ behaving as a resonance, and
$\langle M_1 M_2|S_0\rangle$ indicates that $S_0$ decays into a meson pair.
The $S_0\to M_1 M_2$ strong decay is defined as~\cite{Cheng:2022vbw}
\begin{eqnarray}\label{Amp3a}
\langle M_1 M_2|S_0\rangle\equiv C_{S_0\to M_1 M_2}\,,
\end{eqnarray}
where $C_{S_0\to M_1 M_2}$ is the strong coupling constant.
By denoting $M_0$ as a parity-even scalar meson state,
extending to $S_0$, which may include to-be-mixed light scalar states,
we express the matrix element of the $D\to M_0$ transition 
in a general form~\cite{Becirevic:1999kt,Cheng:2022vbw}:
\begin{eqnarray}\label{Amp3}
&&
\langle M_0|\bar q\gamma_\mu\gamma_5 c|D\rangle=
-i\bigg\{f^+(p^2)(p_D+p_{M_0})_\mu 
+\bigg[f^0(p^2)-f^+(p^2)\bigg]\frac{m^2_D-m^2_{M_0}}{p^2}p_\mu\bigg\}\,,
\end{eqnarray}
where $p_\mu=(p_D-p_{M_0})_\mu$ is the four-momentum energy transfer,
$f^{+,0}(p^2)$ are the form factors. 
Due to parity conservation,
only the matrix elements involving the axial-vector component remain~\cite{Cheng:2022vbw}.
Additionally, given the negligible lepton masses, $f^0(p^2)$ becomes insignificant in the decay,
leaving $f^{+}(p^2)$ as~\cite{BESIII:2023wgr,Becirevic:1999kt},
\begin{eqnarray}\label{t_depent}
f^+(p^2)=\frac{F^{D\to M_0}}{(1-p^2/m^2_{\text A})^2}\,,
\end{eqnarray}
where $m_{\text A}$ is the pole mass. In this equation, 
we define $F^{D\to M_0}\equiv f^+(p^2=0)$, with $p^2=0$ indicating
zero momentum transfer squared. 
This representation captures the momentum dependence of the form factor.

Under $SU(3)_f$ symmetry, the light scalar mesons can have two possible quark structures,
leading to two possible mixing scenarios. For the $q\bar q$ structure, 
the $S_0$ states can be expressed as follows~\cite{Stone:2013eaa,Maiani:2004uc}:
\begin{eqnarray}\label{qqbar}
&&|f_0\rangle= \cos\theta_I|s\bar s\rangle+\sin\theta_I|n\bar n\rangle\,,\nonumber\\
&&|\sigma_0\rangle=-\sin\theta_I|s\bar s\rangle +\cos\theta_I |n\bar n\rangle\,,\nonumber\\
&&
|a_0^0\rangle=|\sqrt {1/2}(u\bar u-d\bar d)\rangle\,,
|a_0^-\rangle=|d\bar u\rangle\,,
\end{eqnarray}
where $|n\bar n\rangle\equiv |\sqrt {1/2}(u\bar u+d\bar d)\rangle$ and
$|s\bar s\rangle$ are mixed as $|f_0\rangle$ and $|\sigma_0\rangle$
with mixing angle $\theta_I$.
Using $|S_0\rangle\sim |q\bar q\rangle$ in Eq.~(\ref{qqbar}),
along with the convenient redefinitions $S_n\equiv |n\bar n\rangle$ and $S_s\equiv |s\bar s\rangle$, 
we can express $\langle S_0(q\bar q),S_n,S_s|\bar q \gamma_\mu\gamma_5 c|D\rangle$ 
as
\begin{eqnarray}\label{mixing1}
&&
\langle f_0,\sigma_0|\bar d \gamma_\mu\gamma_5 c|D^+\rangle
=(\sin\theta_I,\cos\theta_I)\times \langle S_n|\bar d \gamma_\mu\gamma_5 c|D^+\rangle\,,\nonumber\\
&&
\langle a_0^0|\bar d \gamma_\mu\gamma_5 c|D^+\rangle
=\sqrt{1/2}\langle a_0^-|\bar d \gamma_\mu\gamma_5 c|D^0\rangle
=-\langle S_n|\bar d \gamma_\mu\gamma_5 c|D^+\rangle\,,\nonumber\\
&&
\langle f_0,\sigma_0,a_0^0|\bar s \gamma_\mu\gamma_5 c|D^+_s\rangle
=(\cos\theta_I,-\sin\theta_I,0)\times \langle S_s|\bar s \gamma_\mu\gamma_5 c|D^+_s\rangle\,.
\end{eqnarray}
By treating $S_0$ or $S_{n(s)}$ as a parity-even scalar meson $M_0$ and
adopting the definition in Eq.~(\ref{t_depent}), we parameterize
the matrix elements $\langle S_0(q\bar q)|\bar q\gamma_\mu\gamma_5 c|D\rangle$ and
$\langle S_n(S_s)|\bar q\gamma_\mu\gamma_5 c|D^+_{(s)}\rangle$
as the $q\bar q$ structure related form factors at $p^2=0$, 
$F^{D\to S_0}_{q\bar q}$ and $F^{D^+_{(s)}\to S_n(S_s)}$, respectively.
Furthermore, the mixing relations in Eq.~(\ref{mixing1}) give
the following connections:
\begin{eqnarray}\label{FFs1}
&&
(F_{q\bar q}^{D^+\to f_0},F_{q\bar q}^{D^+\to \sigma_0})
=(\sin\theta_I,\cos\theta_I)\times F^{D^+\to S_n}\,,\nonumber\\
&&
F_{q\bar q}^{D^+\to a^0_0}=-F^{D^+\to S_n}\,,\,\nonumber\\
&&
F_{q\bar q}^{D^0\to a^-_0}=-\sqrt 2 F^{D^+\to S_n}\,,\nonumber\\
&&
(F_{q\bar q}^{D_s^+\to f_0},F_{q\bar q}^{D_s^+\to \sigma_0})
=(\cos\theta_I,-\sin\theta_I)\times F^{D_s^+\to S_s}\,.
\end{eqnarray}

Under $SU(3)_f$ symmetry, 
the light scalar mesons have another possible $q^2\bar q^2$ structure~\cite{Maiani:2004uc}:
\begin{eqnarray}\label{qqbarqqbar}
&&|f_0\rangle= \cos\theta_{II}|n\bar n s\bar s \rangle+\sin\theta_{II}|u\bar u d\bar d\rangle\,,\nonumber\\
&&|\sigma_0\rangle=-\sin\theta_{II}|n\bar n s\bar s\rangle +\cos\theta_{II} |u\bar u d\bar d\rangle\,,\nonumber\\
&&
|a_0^0\rangle=|\sqrt {1/2}(u\bar u-d\bar d)s\bar s\rangle\,,
|a_0^-\rangle=|d\bar u s\bar s\rangle\,, 
\end{eqnarray}
where $|n\bar n s\bar s\rangle\equiv |\sqrt {1/2}(u\bar u+d\bar d) s\bar s\rangle$ and
$|u\bar u d\bar d\rangle$ are mixed to form $|f_0\rangle$ and $|\sigma_0\rangle$
with $\theta_{II}$ as the second mixing angle. 
Using $|S_0\rangle\sim |q^2\bar q^2\rangle$ in Eq.~(\ref{qqbarqqbar}) and
redefining $|n\bar n s\bar s\rangle$ as $S_{ns}$ and $|u\bar u d\bar d\rangle$ as $S_{ud}$,
we derive slightly more complicated matrix elements to describe the $D$-meson transitions
to $S_0(q^2\bar q^2)$, $S_{ns}$ and $S_{ud}$,
given by
\begin{eqnarray}\label{mixing2}
&&
\langle f_0|\bar d\gamma_\mu\gamma_5 c|D^+\rangle
=\cos\theta_{II}\langle S_{ns}|\bar d\gamma_\mu\gamma_5 c|D^+\rangle
+\sin\theta_{II}\langle S_{ud}|\bar d\gamma_\mu\gamma_5 c|D^+\rangle\,,
\nonumber\\
&&
\langle \sigma_0|\bar d\gamma_\mu\gamma_5 c|D^+\rangle
=-\sin\theta_{II}\langle S_{ns}|\bar d \gamma_\mu\gamma_5 c|D^+\rangle
+\cos\theta_{II}\langle S_{ud}|\bar d\gamma_\mu\gamma_5 c|D^+\rangle\,,
\nonumber\\
&&
\langle a_0^0|\bar d\gamma_\mu\gamma_5 c|D^+\rangle
=\sqrt{1/2}\langle a_0^-|\bar d\gamma_\mu\gamma_5 c|D^0\rangle
=-\langle S_{ns}|\bar d\gamma_\mu\gamma_5 c|D^+\rangle\,,\nonumber\\
&&
\langle f_0,\sigma_0,a_0^0|\bar s\gamma_\mu\gamma_5 c|D^+_s\rangle
=(\cos\theta_{II},-\sin\theta_{II},0)\times \langle S_{ns}|\bar s\gamma_\mu\gamma_5 c|D^+_s\rangle\,.
\end{eqnarray}
Using Eqs.~(\ref{Amp3})~and~(\ref{t_depent})
and regarding $S_0(q^2\bar q^2)$, $S_{ns}$ or $S_{ud}$ 
as a parity-even scalar meson $M_0$, 
we parameterize $\langle S_0(q^2\bar q^2)|\bar q\gamma_\mu\gamma_5 c|D\rangle$,
$\langle S_{ud}|\bar d\gamma_\mu\gamma_5 c|D^+\rangle$,
$\langle S_{ns}|\bar d\gamma_\mu\gamma_5 c|D^+\rangle$, and
$\langle S_{ns}|\bar s\gamma_\mu\gamma_5 c|D^+_s\rangle$ in Eq.~(\ref{mixing2})
as 
$F^{D\to S_0}_{q^2\bar q^2}$, 
$F^{D^+\to S_{ud}}$,
$F^{D^+\to S_{ns}}$, and
$F^{D^+_{s}\to S_{ns}}$, 
respectively, which are the remaining $q^2\bar q^2$ structure-related form factors at $p^2=0$,
as defined by $F^{D\to M_0}$ in Eq.~(\ref{t_depent}). 
Furthermore, we present the mixtures of the form factors as
\begin{eqnarray}\label{FFs2}
&&
F^{D^+\to f_0}_{q^2\bar q^2}=
\cos\theta_{II}F^{D^+\to S_{ns}}+\sin\theta_{II}F^{D^+\to S_{ud}}\,,\nonumber\\
&&
F^{D^+\to \sigma_0}_{q^2\bar q^2}=
-\sin\theta_{II}F^{D^+\to S_{ns}}+\cos\theta_{II}F^{D^+\to S_{ud}}\,,\nonumber\\
&&
F^{D^+\to a_0^0}_{q^2\bar q^2}=-F^{D^+\to S_{ns}}\,,\,\nonumber\\
&&
F^{D^0\to a_0^-}_{q^2\bar q^2}=-\sqrt 2 F^{D^+\to S_{ns}}\,,\nonumber\\
&&
(F^{D_s^+\to f_0}_{q^2\bar q^2},F^{D_s^+\to \sigma_0}_{q^2\bar q^2})
=(\cos\theta_{II},-\sin\theta_{II})\times F^{D_s^+\to S_{ns}}\,.
\end{eqnarray}
Additionally, $S_0(q^2\bar q^2)$ has an extra $q\bar q$ pair
compared with $S_0(q\bar q)$, requiring $g\to s\bar s(u\bar u)$
in the $D^+\to d\bar d\to S_{ns(ud)}$ transition,
where $g$ denotes a gluon originating from the vacuum.  
Since it is possible that under the $SU(3)_f$ symmetry
$D^+\to d\bar d\to S_{ud}$ with $g\to u\bar u$ is not distinguished from
$D^+\to d\bar d\to S_{ns}$ with $g\to s\bar s$,
we assume $F^{D^+\to S_{ud}}=\sqrt 2 F^{D^+\to S_{ns}}$.

In Eq.~(\ref{Amp2}), the scalar meson as a resonance of 
$D\to S_0 e^+ \nu_e,S_0\to M_1 M_2$ is presented by $D^{-1}_{S_0}$.
Due to the multiple decay channels of $f_0\to(\pi\pi,K\bar K)$ or $a_0\to(\eta\pi, K\bar K)$, 
the Flatte formula has been extensively applied~\cite{Flatte:1976xu,Flatte:1976xv,Baru:2004xg}. 
Since $\Gamma_{\sigma_0}\sim m_{\sigma_0}$,  it has been stressed that 
$D^{-1}_{\sigma_0}$ cannot be expressed with the usual Breit-Wigner formula~\cite{Pelaez:2015qba},
where the constant decay width is only appropriate for a narrow resonance~\cite{pdg}. Therefore,
one adopts the energy-dependent-width Breit-Wigner parameterization~\cite{pdg,Ceci:2013zta},
which is specifically applied to a resonance with a decay width as broad as its mass.

The expressions of $D_{f_0}$, $D_{a_0}$, and $D_{\sigma_0}$
can be found in Refs.~\cite{Flatte:1976xv,Bugg:2008ig,LHCb:2014ooi,LHCb:2014vbo}, 
\cite{Abele:1998qd,BESIII:2020hfw}, and~\cite{LHCb:2019sus,Oller:2004xm,Cheng:2022vbw,
BES:2004mws,Zou:1993az}, respectively:
\begin{eqnarray}\label{RS0}
&&
D_{f_0}=m_{f_0}^2-t-im_{f_0}(g_{f_0 \pi\pi}\rho_{\pi\pi}+g_{f_0 KK}\rho_{KK}F_{KK}^2)\,,\nonumber\\
&&
D_{a_0}=m_{a_0}^2-t-i(g_{a_0\pi\eta}^2\rho_{\pi\eta}+g_{a_0 KK}^2\rho_{KK})\,,\nonumber\\
&&
D_{\sigma_0}={m_{\sigma_0}^2-t-\Gamma_{\sigma_0}^2(t)/4-i\,m_{\sigma_0}\Gamma_{\sigma_0}(t)}\,,
\end{eqnarray}
where $t\equiv (p_{M_1}+p_{M_2})^2$, and $\rho_{M_1 M_2}=[(1-m_+^2/t)(1-m_-^2/t)]^{1/2}$ 
with $m_{\pm}=m_{M_1}\pm m_{M_2}$. In Eq.~(\ref{RS0}), 
$D_{\sigma_0}$ accounts for the pole position 
of $\sqrt{s_{\sigma_0}}=m_{\sigma_0}-i\Gamma_{\sigma_0}/2$~\cite{Pelaez:2015qba};
additionally, the decay width is energy dependent:
$\Gamma_{\sigma_0}(t)\equiv (\rho_{\pi\pi}/\bar\rho_{\pi\pi})\Gamma^0_{\sigma_0}$,
where $\bar\rho_{\pi\pi}$ denotes $\rho_{\pi\pi}$ at $t=m_{\sigma_0}^2$. 
For $D^{-1}_{S_0}$, the resonant parameters: the resonant masses,
the strong coupling constants, and the decay width, 
have been extracted in Refs.~\cite{Flatte:1976xu,Flatte:1976xv,
Baru:2004xg,Abele:1998qd,Bugg:2008ig,LHCb:2014ooi,LHCb:2014vbo} 
for reproducing the observed line shapes of the $\pi\pi$ invariant mass spectrums.
Particularly, the factor $F_{KK}=e^{-\alpha k^2}$, 
with $\alpha=2$~GeV$^{-2}$~\cite{LHCb:2014ooi,LHCb:2014vbo}
and $k$ the momentum of each kaon in the $KK$ rest frame,
is introduced to reduce $\rho_{KK}$ above the $KK$ threshold~\cite{Bugg:2008ig},
avoiding the unreasonable phase space $\rho_{KK}\sim 1$ at $t\to \infty$.

%
\begin{figure}[t!]
\centering
\includegraphics[width=3.8in]{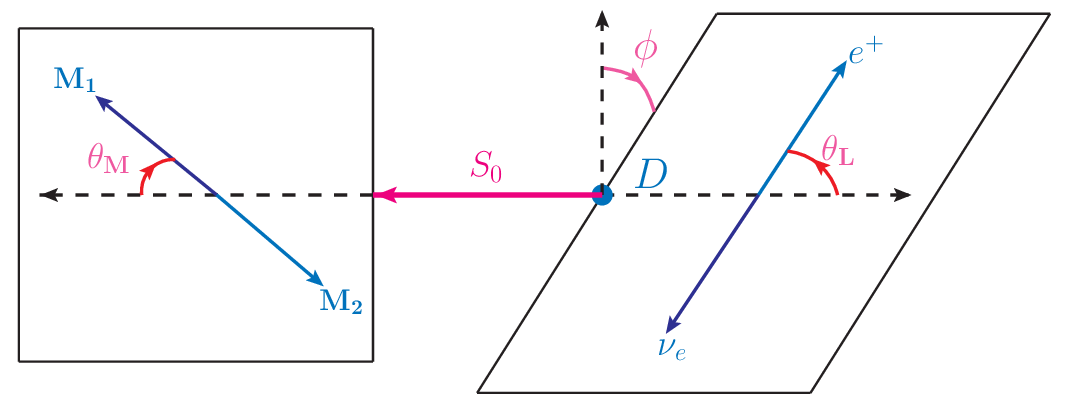}
\caption{$\theta_{\bf M}$, $\theta_{\bf L}$ and $\phi$:
the angular variables of the four-body $D\to S_0 e^+ \nu_e,S_0\to M_1 M_2$ decay.}\label{fig1}
\end{figure}
%

In the resonant four-body 
$D(p_D)\to S_0(p_{S_0}) e^+(p_e)\nu(p_\nu),S_0(p_{S_0})\to M_1(p_1) M_2(p_2) $ decay,
one has $s\equiv (p_e+p_\nu)^2\equiv m_{e\nu}^2$, $t$, 
and the angular variables $(\theta_{\bf M},\theta_{\bf L},\phi)$ in the phase space~\cite{Kl4,Wise,Tsai:2021ota}.
We illustrate the angular variables in Fig.~\ref{fig1}, where
$\theta_{\bf M(L)}$ is the angle between $\vec{p}_1$  ($\vec{p}_e$) in the $M_1 M_2$ ($e\nu$) rest frame
and the resonant $S_0$ moving direction (the line of flight of $e\nu$ system) in the $D$ meson rest frame,
while $\phi$ is the angle between the $M_1 M_2$ and $e\nu$ planes,
formed by the momenta of the $M_1 M_2$ and $e\nu$ systems, respectively, in the $D$ meson rest frame.
The partial decay width thus reads~\cite{Hsiao:2022tfj,Hsiao:2022uzx}
\begin{eqnarray}\label{Gamma1}
d\Gamma=\frac{|\bar {\cal M}|^2}{4(4\pi)^6 m_D^3}X(s,t)
\alpha_{\bf M}\alpha_{\bf L}\, ds\, dt\, d\text{cos}\,\theta_{\bf M}\, d\text{cos}\,\theta_{\bf L}\, d\phi\,,
\end{eqnarray}
where 
$X(s,t)=[(m_D^2-s-t)^2/4-st]^{1/2}$,
$\alpha_{\bf M}=\lambda^{1/2}(t,m_1^2,m_2^2)/t$, and
$\alpha_{\bf L}=\lambda^{1/2}(s,m_e^2,m_\nu^2)/s$,
with $\lambda(a,b,c)=a^2+b^2+c^2-2ab-2bc-2ca$. 
The allowed regions of the variables are:
\begin{eqnarray}
&&
(m_e+m_\nu)^2\leq s\leq (m_D-\sqrt{t})^2\,,\nonumber\\
&&
(m_1+m_2)^2\leq t\leq (m_D-m_e-m_\nu)^2\,,\nonumber\\
&&
0\leq \theta_{\bf M,L}\leq \pi\,,\; 0\leq \phi\leq 2\pi\,.
\end{eqnarray}

\section{Numerical Results}
In the numerical analysis, we use $V_{cs}=1-\lambda^2$ and $V_{cd}=-\lambda$
with $\lambda=0.22453\pm 0.00044$ in the Wolfenstein parameterization~\cite{pdg}. 
The mixing angle $\theta_I$ in the $q\bar q$ constituent quark model 
has been extensively studied in
$f_0\to (KK,\pi\pi)$, $(f_0,a_0)\to \gamma\gamma$, $J/\psi\to (\sigma_0,f_0)\gamma$,
$D^{+(0)}\to \pi^{+(0)}\pi^+\pi^-$, $\bar B^0\to D^0\pi^+\pi^-$, 
which lead to different extractions, as listed in Table~\ref{thetaI}.
Therefore, we compute a weighted average of the extracted numbers 
using the standard procedure, given by~\cite{pdg}
\begin{eqnarray}\label{WAthetaI}
\bar x\pm \delta x=\frac{\Sigma_i w_i x_i}{\Sigma_i w_i}\pm \bigg(\Sigma_i w_i\bigg)^{-1/2}\,,
\end{eqnarray}
where $w_i=1/\sigma_i^2$. 
%
\begin{table}[b]
\caption{Observables and the mixing angle $\theta_I$ for $q\bar q$ determined from them, where
${\cal R}_{D_1}\equiv{\Gamma(D_s^+\to f_0\pi^+)}/{\Gamma(D^+\to f_0\pi^+)}$ and
${\cal R}_{D_2}\equiv{\Gamma(D_s^+\to f_0\pi^+)}/{\Gamma(D^+\to K^*_0(1430)\pi^+)}$.
}\label{thetaI}
{
\footnotesize
\begin{tabular}{lc}
\hline
observable
&$\theta_I$\\
\hline\hline
${\cal B}(f_0\to KK,\pi\pi)$
&$(162\pm 1)^\circ$~\cite{Ochs:2013gi}\\
${\cal B}(\bar B_s^0\to J/\psi f_0)$
&$(146^{+15}_{-\;\,9})^\circ$~\cite{Li:2012sw}\\
${\cal B}(\bar B^0\to D^0\pi^+\pi^-)$
&$(157\pm 4)^\circ$~\cite{LHCb:2015klp}\\
${\cal B}(\phi\to f_0\gamma)$
&$(138\pm 6)^\circ$~\cite{Anisovich:2002wy}\\
$\frac{{\cal B}(J/\psi\to f_0\phi)}{{\cal B}(J/\psi\to f_0\omega)}$
&$(147\pm 7)^\circ$~\cite{DiDonato:2011kr,Ochs:2013gi}\\
$\frac{\Gamma(f_0\to \gamma\gamma)}{\Gamma(a_0\to \gamma\gamma)}$
&$(126\pm 8)^\circ$~\cite{Minkowski:1998mf,Ochs:2013gi}\\
${\cal R}_{D_1}$, ${\cal R}_{D_2}$
&$(161^{+26}_{-17})^\circ$~\cite{Ochs:2013gi}\\
${\cal B}(J/\psi\to \sigma_0\gamma,f_0\gamma)$
&$(153\pm 1)^\circ$~\cite{Sarantsev:2021ein,Klempt:2021nuf}\\
${\cal B}(D^+\to \pi^+\pi^-\pi^+)$
&$(157\pm 3)^\circ$~\cite{Klempt:2021nuf}\\
${\cal B}(D^0\to \pi^0\pi^+\pi^-)$
&$(145\pm 5)^\circ$~\cite{Klempt:2021nuf}\\
\hline
\end{tabular}}
\end{table}
%
As we calculate the weighted average of the values for $\theta_{I}$ 
from different studies~\cite{Ochs:2013gi,Li:2012sw,
LHCb:2015klp,Anisovich:2002wy,DiDonato:2011kr,Ochs:2013gi,Minkowski:1998mf,
Sarantsev:2021ein,Klempt:2021nuf,Braghin:2022uih},
$\theta_{II}$ can be found in~\cite{Maiani:2004uc}.
We thus present the mixing angles in the $q\bar q$ and $q^2\bar q^2$ pictures as
\begin{eqnarray}\label{theta12}
\theta_{I}=(156.7\pm 0.7)^\circ\,,\;
\theta_{II}=(174.6^{+3.4}_{-3.2})^\circ\,.
\end{eqnarray}
Following Refs.~\cite{Becirevic:1999kt,pdg}, we use
$m_{\text A}=(2459.5\pm 0.6,2422.1\pm 0.6)$~MeV~\cite{pdg}
in the $D_s^+$ and $D^{+,0}$ decays, respectively.

For the resonant parameters of $f_0$, $a_0$, and $\sigma_0$ in Eq.~(\ref{RS0}),
we adopt the experimentally extracted values from~\cite{LHCb:2014ooi,LHCb:2014vbo}, 
\cite{Abele:1998qd,BaBar:2005vhe}, 
and~\cite{Sarantsev:2021ein,LHCb:2019sus,BESIII:2016tdb},
respectively, as the inputs in the global fit, 
instead of using the theoretical calculations,
which may cause considerable uncertainties 
in determining the form factors. Here, we summarize 
the resonant parameters in this extraction:
\begin{eqnarray}\label{coupling1}
&&
(m_{f_0},g_{f_0 \pi\pi})=(949.9\pm 2.1,167\pm 8)~\text{MeV}\,,\;
g_{f_0 KK}=(3.05\pm 0.13)g_{f_0 \pi\pi}\,,\nonumber\\
&&
(m_{a_0},g_{a_0 \pi\eta})=(999\pm 2,324\pm 15)~\text{MeV}\,,\;
g^2_{a_0 KK}=(1.03\pm 0.14)g^2_{a_0 \pi\eta}\,,\nonumber\\
&&
(m_{\sigma_0},\Gamma_{\sigma_0}^0)=
(527.0\pm 7.7,513.5\pm 15.2)~\text{MeV}\,,
\end{eqnarray}
where $m_{\sigma_0}$ and $\Gamma_{\sigma_0}^0$ are
the weighted averaged values of \cite{Sarantsev:2021ein,LHCb:2019sus,BESIII:2016tdb} 
for the most current experimental inputs. 
In Eq.~(\ref{Amp3a}),
$C_{S_0\to M_1 M_2}$ is a variant form of $g_{S_0 M_1 M_2}$,
related as~\cite{Baru:2004xg,Bugg:2008ig}
\begin{eqnarray}\label{coupling2}
&&
C_{f_0\to\pi^+\pi^-}=\sqrt 2 C_{f_0\to\pi^0\pi^0}
=(8\pi m_{f_0}^2 g_{f_0 \pi\pi})^{1/2}\,,\nonumber\\
&&
C_{a_0^{-(0)}\to\pi^{-(0)} \eta}=4\sqrt{\pi}g_{a_0 \pi\eta}\,,\nonumber\\
&&
C_{\sigma_0\to\pi^+\pi^-}=\sqrt 2 C_{\sigma_0\to\pi^0\pi^0}=
[(8\pi m_{\sigma_0}^2/|\vec{p}_{\rm cm}|)(2/3)\Gamma_{\sigma_0}^0]^{1/2}\,,
\end{eqnarray}
where $C_{\sigma_0\to\pi^+\pi^-}$
is obtained by utilizing the decay-width equation~\cite{Cheng:2022vbw}.
Consequently, no additional parameters are introduced 
for the resonant $S_0\to M_1 M_2$ decays. 

We perform a statistical global fit to extract the  $D$ to $S_0$ transition form factors.
Using the minimum $\chi^2$-fit equation,
\begin{eqnarray}\label{chi2}
\chi^2=\sum_i\frac{(\hat {\cal B}_i-{\cal B}_i)^2}{(\sigma_{{\cal B}_i})^2}
+\frac{(\hat\theta_k-\theta_k)^2}{(\sigma_{\theta_k})^2}\,,
\end{eqnarray}
the global fit is carried out in two scenarios: 
one for the $q\bar q$ structure and the other for the $q^2\bar q^2$ structure.
In the above equation, ${\cal B}_i$ and $\theta_j$ with $j=(I,II)$ represent
the measured ${\cal B}(D\to  S_0 e^+\nu_e,S_0\to M_1 M_2)$ in Table~\ref{pre}
and the mixing angles in Eq.~(\ref{theta12}), respectively.
The terms $(\sigma_{{\cal B}_i},\sigma_{\theta_j})$ denote the corresponding errors,
while $(\hat {\cal B}_i,\hat\theta_j)$ with the hat notions refer to the fitted items.
The parameters $(F^{D^+\to S_n}, F^{D_s^+\to S_s})$ for the $q\bar q$ scenario 
and $(F^{D^+\to S_{ns}}, F^{D_s^+\to S_{ns}})$ for the $q^2\bar q^2$ scenario 
are treated as free parameters in the global fit.
%
\begin{figure}[t!]
\centering
\includegraphics[width=2in]{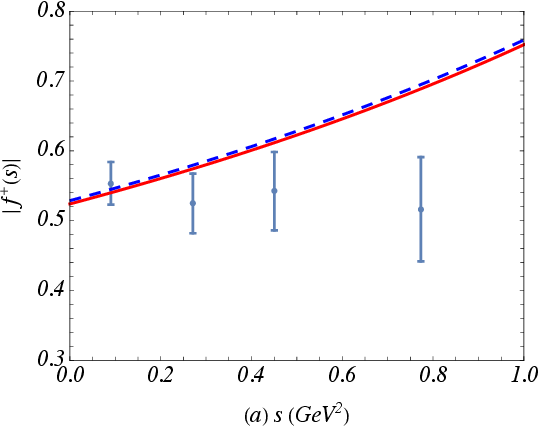}
\includegraphics[width=2in]{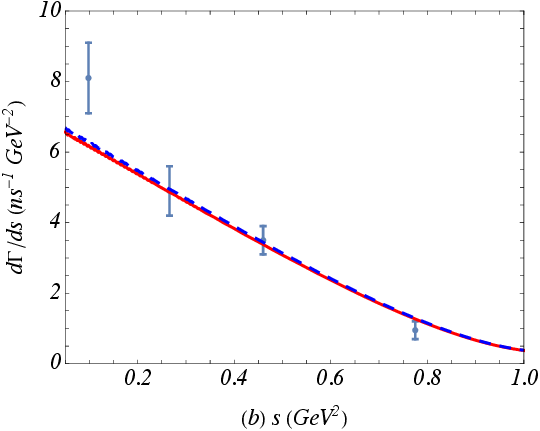}
\caption{(a) The form factor of the $D_s^+\to f_0$ transition $|f^+(s)|$ 
and (b) the differential decay width of $D_s^+\to f_0 e^+\nu_e, f_0\to\pi^+\pi^-$
are both shown as functions of $s$. The red solid and blue dashed lines correspond to
the $q\bar q$ and $q^2\bar q^2$ structures, respectively, 
with data points taken from Ref.~\cite{BESIII:2023wgr}.}\label{fig2}
\end{figure}
%
%
\begin{table}[b!]
\caption{Form factors of our work, resulting from the $SU(3)_f$ relations in Eqs.~(\ref{FFs1}, \ref{FFs2}) 
and the extractions in Eqs.~(\ref{fitFFs1}, \ref{fitFFs2}), 
in comparison with the existing model calculations and extractions.}\label{FFs_result}
{
\tiny
\begin{tabular}{lccccc}
\hline
Form factor &$|F^{D_s^+\to f_0}|$&$|F^{D_s^+\to \sigma_0}|$
&$|F^{D^0\to a_0^-}|$&$|F^{D^+\to f_0}|$&$|F^{D^+\to \sigma_0}|$\\
\hline\hline
our work: $(q\bar q)$ 
& $0.52\pm 0.02$
& $0.22\pm 0.01$
& $0.66\pm 0.03$
& $0.19\pm 0.01$
&$0.43\pm 0.02$\\
our work: $(q^2\bar q^2)$ 
& $0.53\pm 0.02$
& $0.037\pm 0.032$
& $0.44\pm 0.03$
& $0.28\pm 0.03$
&$0.46\pm 0.03$\\
\hline
Fit to ${\cal B}(D\to SP)$~\cite{Cheng:2002ai}
&$0.52\pm 0.04$
&
&
& $0.26\pm 0.02$
&$0.42\pm 0.05$
\\
CQM~\cite{Gatto:2000hj}
&
&
&
&
&$0.57\pm 0.09$
\\
Fit to E791 data~\cite{Boito:2008zk}
&
&
&
&
&$0.38\pm 0.06$
\\
%
CLFD (DR)~\cite{El-Bennich:2008rkp}
&0.45 (0.46)
&
&
&0.21 (0.22)
&
\\
QCDSR~\cite{Bediaga:2003zh}
&$0.50\pm 0.13$
&
&
&$0.53\pm 0.15$
&
\\
QCDSR~\cite{Aliev:2007uu}
&$0.46\pm 0.03$
&
&
&$0.54\pm 0.05$
&
\\
&$(0.27\pm 0.02)$
&
&
&$(0.32\pm 0.03)$
&
\\
%
LFQM~\cite{Ke:2009ed}
&0.43
&
&
&0.22
&
\\
%
CCQM~\cite{Soni:2020sgn}
&$0.39\pm 0.02$
&
&$0.55\pm 0.02$
&$0.45\pm 0.02$
&
\\
%
LCSR~\cite{Colangelo:2010bg}
&$0.30\pm 0.03$
&
&
&
&
\\
%
LCSR~\cite{Cheng:2017fkw}
&
&
&$0.88\pm 0.14$
&
&
\\
%
LCSR~\cite{Huang:2021owr}
&
&
&$0.85^{+0.10}_{-0.11}$
&
&
\\
%
LCSR~\cite{Wu:2022qqx}
&
&
&$1.070^{+0.066}_{-0.033}$
&
&
\\
%
AdS/QCD~\cite{Momeni:2022gqb}
&
&
&$0.72\pm 0.09$
&
&
\\
\hline
\end{tabular}}
\end{table}
%

Subsequently, we determine
\begin{eqnarray}\label{fitFFs1}
F^{D^+\to S_n}=0.47\pm 0.02\,,\; 
F^{D_s^+\to S_s}=0.57\pm 0.02\,, 
\end{eqnarray}
with $\chi^2_{q\bar q}/n.d.f\simeq 1.9$ for the $q\bar q$ structure, where $n.d.f=3$ 
represents the number of degrees of freedom.
For the $q^2\bar q^2$ structure, we determine
\begin{eqnarray}\label{fitFFs2}
F^{D^+\to S_{ns}}=0.31\pm 0.02\,,\;
F^{D_s^+\to S_{ns}}=0.53\pm 0.02\,,
\end{eqnarray}
with $\chi^2_{q^2\bar q^2}/n.d.f\simeq 1.4$. Due to 
$F^{D^+\to S_{ud}}=\sqrt 2 F^{D^+\to S_{ns}}$,
we get $F^{D^+\to S_{ud}}=0.44\pm 0.02$. We thus obtain 
$F^{D\to S_0}_{(q\bar q, q^2\bar q^2)}$ in Table~\ref{FFs_result}, 
by which we calculate the branching fractions in Table~\ref{pre}.
Additionally, we depict the form factor $|f^+(s)|$ for the $D_s^+\to f_0$ transition 
in Fig.~\ref{fig2}a and the differential decay width of $D_s^+\to f_0 e^+\nu_e, f_0\to\pi^+\pi^-$
in Fig.~\ref{fig2}b, and the partial branching fraction of $D_s^+\to f_0 e^+\nu_e, f_0\to\pi^+\pi^-$
in Fig.~\ref{fig3} to be compared to the experimental data~\cite{BESIII:2023wgr}.

\section{Discussion and Conclusions}
%
\begin{figure}[t!]
\centering
\includegraphics[width=2.4in]{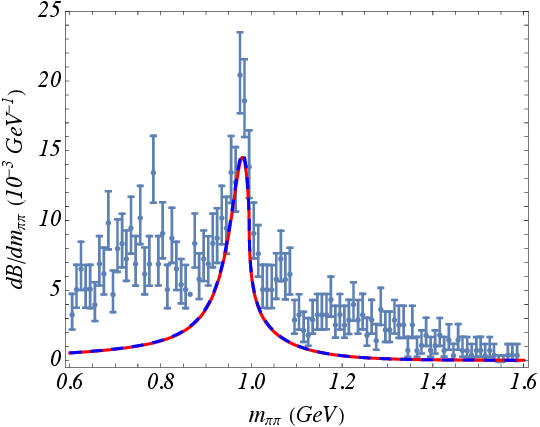}
\caption{
The $\pi\pi$ invariant mass spectrum for $D_s^+\to f_0 e^+\nu_e, f_0\to\pi^+\pi^-$
is shown.
The red solid and blue dashed lines correspond to
the $q\bar q$ and $q^2\bar q^2$ structures, respectively, 
with data points taken from Ref.~\cite{BESIII:2023wgr}.}\label{fig3}
\end{figure}
%
%
\begin{table}[b]
\caption{Branching fractions of the resonant $D\to S_0e^+\nu_e,S_0\to M_1 M_2$ decays
in the $q\bar q$ and $q^2\bar q^2$ structures, in comparison with the experimental data.
The first errors of our work consider the uncertainties of the form factors 
in Table~\ref{FFs_result}, and the second one combines 
those of the resonant parameters in Eq.~(\ref{coupling1}).
The superscript $\dagger$ denotes the number in the isospin relation.}\label{pre}
{
\tiny
\begin{tabular}{lll}
\hline
decay channel&our work:~$(q\bar q, q^2\bar q^2)$&experimental data\\
\hline\hline
$10^4 {\cal B}(D_s^+\to  f_0 e^+\nu_e,f_0\to\pi^+\pi^-)$
& ($16.9\pm 1.2\pm 0.7$, $17.2\pm 1.3\pm 0.7$) 
&$17.1\pm 1.6$~\cite{BESIII:2023wgr}\\
$10^4 {\cal B}(D_s^+\to  f_0 e^+\nu_e,f_0\to\pi^0\pi^0)$
&($8.4\pm 0.6\pm 0.4$, $8.6\pm 0.7\pm 0.4$)
&$7.9\pm 1.4$~\text{\cite{BESIII:2021drk}}\\
$10^4 {\cal B}(D_s^+\to  \sigma_0 e^+\nu_e,\sigma_0\to\pi^+\pi^-)$
&($20.3\pm 1.8\pm 0.5$, $0.58^{+1.43}_{-0.57}\pm 0.01$)
&$<3.3$~\cite{BESIII:2023wgr}\\
$10^4 {\cal B}(D_s^+\to  \sigma_0 e^+\nu_e,\sigma_0\to\pi^0\pi^0)$
&($10.7\pm 1.0\pm 0.3$, $0.29^{+0.72}_{-0.29}\pm 0.01$)
&$<7.3$~\text{\cite{BESIII:2021drk}}~($<1.7$~\cite{BESIII:2023wgr})$^\dagger$\\
\hline
$10^4 {\cal B}(D^0\to  a_0^- e^+\nu_e,a_0^-\to\pi^-\eta)$
&($1.8\pm 0.2\pm 0.2$, $0.8\pm 0.1\pm 0.1$)
&$1.3\pm 0.3$~\text{\cite{pdg}}\\
$10^4 {\cal B}(D^+\to  a_0^0 e^+\nu_e,a_0^0\to\pi^0\eta)$
&($2.4\pm 0.2\pm 0.3$, $1.0\pm 0.1\pm 0.1$)
&$1.7\pm 0.7$~\text{\cite{pdg}}\\
$10^5 {\cal B}(D^+\to  f_0 e^+\nu_e,f_0\to\pi^+\pi^-)$
&$(1.3\pm 0.1\pm 0.1,2.9\pm 0.7\pm 0.2)$
&$<2.8$~\text{\cite{BESIII:2018qmf}}\\
$10^4 {\cal B}(D^+\to  \sigma_0 e^+\nu_e,\sigma_0\to\pi^+\pi^-)$
&($5.4\pm 0.5\pm 0.1$, $6.2\pm 0.9\pm 0.1$)
&$6.3\pm 0.5$~\text{\cite{BESIII:2018qmf}}\\
\hline
\end{tabular}}
\end{table}
%
Our approach carefully considers the $SU(3)_f$ relations of the form factors
in the $q\bar q$ and $q^2\bar q^2$ pictures, together with the momentum dependence,
the mixtures, the mixing angles, and the resonant effects.
Since the $\chi^2_{(q\bar q,q^2\bar q^2)}/n.d.f\simeq (1.9,1.4)$
indicate the goodness of the global fit, the feasibility of our approach is established.
Additionally, the form factors $|F^{D\to S_0}_{(q\bar q,q^2\bar q^2)}|$ in Table~\ref{FFs_result} 
as well as the resulting first errors of ${\cal B}(D\to S_0 e^+\nu_e,S_0\to M_1 M_2)$  in Table~\ref{pre}
demonstrate the well-controlled uncertainties.
This reflects the fact that the $\chi^2$-fit has used 
the currently highest-precision experimental results as inputs.

As shown in the theoretical predictions in Table~\ref{FFs_result},
$|F^{D_s^+\to f_0}_{(q\bar q,q^2\bar q^2)}|=(0.52\pm 0.02,0.53\pm 0.02)$,
which are in excellent agreement with the experimental value of $0.52\pm 0.04$ 
obtained by BESIII~\cite{BESIII:2023wgr}. Additionally,
we depict $f^+(s)$ and the differential decay width of $D_s^+\to f_0 e^+\nu_e,f_0\to\pi^+\pi^-$,
both as functions of $s$, which agree with the four data points
in Fig.~\ref{fig2}a and Fig.~\ref{fig2}b, respectively. In particular, 
the resulting two lines for the $q\bar q$ and $q^2\bar q^2$ structures
are as expected to be close to each other. Consequently,
our determinations for the form factors are justified.

We incorporate the resonant expressions in Eq.~(\ref{RS0}) 
to explore the resonant decay $D\to S_0 e^+\nu, S_0\to M_1 M_2$, instead of
using the approximation ${\cal B}(D\to S_0 e^+\nu, S_0\to M_1 M_2)
\simeq {\cal B}(D\to S_0 e^+\nu)\times {\cal B}(S_0\to M_1 M_2)$, 
which is only appropriate for the usual Breit-Wigner resonance with a narrow width~\cite{pdg}.
Thus, we ensure that the resonant effects due to $f_0$, $a_0$, and $\sigma_0$
in this study can be as accurate as the experimental studies
in~\cite{LHCb:2014ooi,LHCb:2014vbo}, \cite{Abele:1998qd,BaBar:2005vhe} 
and~\cite{Sarantsev:2021ein,LHCb:2019sus}, respectively, 
where the observed line shapes in the $\pi\pi$ invariant mass spectrums
have been reproduced using the same resonant parameters as those in Eq.~(\ref{coupling1}).
In Fig.~\ref{fig3}, 
the $\pi\pi$ invariant mass spectrum of $D_s^+\to f_0 e^+\nu_e, f_0\to \pi^+\pi^-$
demonstrates that the lines, accounting for the resonant effect, interpret the data well.

Experiments such as BESIII combine the uncertainties of
the resonant parameters as the second errors of ${\cal B}(D\to S_0 e\nu,S_0\to M_1 M_2)$.
Likewise, we estimate the second errors of ${\cal B}(D\to S_0 e\nu,S_0\to M_1 M_2)$
due to the uncertainties of the resonant parameters in Eq.~(\ref{coupling1}). 
Evidently, the smallness of
$\delta m_{a_0}\simeq 0.002 m_{a_0}$, $\delta m_{f_0}\simeq 0.002 m_{f_0}$, and
$\delta m_{\sigma_0}\simeq 0.015 m_{\sigma_0}$ cause insignificant errors. 
Although $\delta g_{f_0 \pi\pi}=0.05g_{f_0 \pi\pi}$, $\delta g_{a_0 \pi\eta}=0.05g_{a_0 \pi\eta}$, 
and $\delta\Gamma_{\sigma_0}^0=0.03 \Gamma_{\sigma_0}^0$ in $D_{S_0}^{-1}$
might cause a few percents of errors, we observe that
each $C_{S_0\to M_1 M_2}$ due to the relation in Eq.~(\ref{coupling2})
has a correlation with its own $g_{S_0 M_1 M_2}$ or $\Gamma_{S_0}^0$,
thereby reducing the uncertainties caused by $D_{S_0}^{-1}$. Consequently,
the remaining uncertainties, such as $\delta g_{f_0 KK}\simeq 0.04 g_{f_0 \pi\pi}$ 
and $\delta g_{a_0 KK}^2\simeq 0.136 g_{a_0 \eta\pi}^2$, are responsible for
most of the second errors. Based on the well-controlled second errors,
we demonstrate that the resonant expressions in Eq.~(\ref{RS0}), together with
the resonant parameters in Eq.~(\ref{coupling1}), 
in no way compromise our numerical results.

Our results are characterized as being independent of QCD or quark model calculations
due to the incorporation of both the $SU(3)_f$ relation and data extraction.
As a result, we present
$F^{D\to S_0}_{(q\bar q, q^2\bar q^2)}$ for $S_0 = (a_0, f_0, \sigma_0)$
for the first time, which provide a comprehensive comparison 
to the model-dependent calculations shown in Table~\ref{FFs_result}. 
It is interesting to note that the theoretical calculations have not provided
all of the form factors in the $q\bar q$ structure; particularly,
those of the $q^2\bar q^2$ structure are currently unavailable.
Hence, our form factors can be informative.

Utilizing $|F^{D_s^+\to\sigma_0}_{(q\bar q,q^2\bar q^2)}|=(0.22\pm 0.01,0.037\pm 0.032)$
and considering the resonant effect,
we calculate the branching fractions 
${\cal B}_{(q\bar q,q^2\bar q^2)}^{D\to S_0\to M_1 M_2}\equiv
{\cal B}_{(q\bar q,q^2\bar q^2)}(D\to S_0 e^+\nu_e,S_0\to M_1 M_2)$
as
\begin{eqnarray}\label{2results1}
{\cal B}_{(q\bar q,q^2\bar q^2)}^{D_s^+\to \sigma_0\to \pi^+\pi^-}
&=&(20.3\pm 1.8\pm 0.5,0.58^{+1.43}_{-0.57}\pm 0.01)\times 10^{-4}\,,
\nonumber\\
{\cal B}_{(q\bar q,q^2\bar q^2)}^{D_s^+\to \sigma_0\to \pi^0\pi^0}
&=&(10.7\pm 1.0\pm 0.3,0.29^{+0.72}_{-0.29}\pm 0.01)\times 10^{-4}\,,
\end{eqnarray}
where
${\cal B}^{D_s^+\to \sigma_0\to \pi^+\pi^-}\simeq 2{\cal B}^{D_s^+\to \sigma_0\to \pi^0\pi^0}$
manifests the isospin symmetry. It is resonable that
${\cal B}_{\rm ex}^{D_s^+\to \sigma_0\to \pi^+\pi^-}<3.3\times 10^{-4}$~\cite{BESIII:2023wgr}
can lead to ${\cal B}_{\rm exp}^{D_s^+\to \sigma_0\to \pi^0\pi^0}<1.7\times 10^{-4}$,
which is more stringent than the number of $7.3\times 10^{-4}$ in~\cite{BESIII:2021drk},
due to the same isospin relation.
 
Clearly, the light scalar meson as a normal $q\bar q$ meson can be excluded, 
given that ${\cal B}_{(q\bar q)}^{D_s^+\to \sigma_0\to \pi^{+(0)}\pi^{-(0)}}$ significantly
causes $9\sigma$ ($10\sigma$) deviations deviating from the upper limit of 
${\cal B}_{\rm ex}^{D_s^+\to \sigma_0\pi^{+(0)}\pi^{-(0)}}$. 
In contrast, ${\cal B}_{(q^2\bar q^2)}^{D_s^+\to \sigma_0\to \pi^{+(0)}\pi^{-(0)}}$ 
is within the experimental allowed range. This can be attributed to
$\sin\theta_{II}\sim 0$, causing the slightest mixture of $|n\bar n s\bar s\rangle$
for the $\sigma_0$ meson. 
As a result, we test the light scalar meson as a non-$q\bar q$ state. 

The semileptonic $D^{+(0)}$ decays are not as informative as the $D_s^+$ ones for exploring
the light scalar meson. The primary reason is that the discrepancy between
$|F_{q\bar q}^{D^{+(0)}\to S_0}|$ and $|F_{q^2\bar q^2}^{D^{+(0)}\to S_0}|$
is not as large as that between 
$|F_{q\bar q}^{D^+_s\to \sigma_0}|$ and $|F_{q^2\bar q^2}^{D^+_s\to \sigma_0}|$.
Consequently, we cannot favor either quark content in $D^{+(0)}\to S_0 e\nu,S_0\to M_1 M_2$, 
as either ${\cal B}_{q\bar q}$ or ${\cal B}_{q^2\bar q^2}$ is calculated to match ${\cal B}_{\rm ex}$. 
However, if $\delta{\cal B}_{\rm ex}(D^{0(+)}\to  a_0^{-(0)}\to\pi^{-(0)}\eta)$ 
is reduced in future measurements, exploration might become feasible,
given that ${\cal B}_{q\bar q}^{D^{0(+)}\to  a_0^{-(0)}\to\pi^{-(0)}\eta}$
is twice as large as ${\cal B}_{q^2\bar q^2}^{D^{0(+)}\to  a_0^{-(0)}\to\pi^{-(0)}\eta}$. 
Indeed, $F_{q^2\bar q^2}^{D^+\to  \sigma_0}$ and $F_{q^2\bar q^2}^{D^+\to  f_0}$
are the mixtures of $F^{D^+\to F_{ns}}$ and $F^{D^+\to F_{ud}}$,
constrained by ${\cal B}_{\rm ex}^{D^+\to\sigma_0\to\pi^+\pi^-}$
and ${\cal B}_{\rm ex}^{D^+\to f_0\to\pi^+\pi^-}$. 
Since ${\cal B}_{\rm ex}^{D^+\to f_0\to\pi^+\pi^-}$ is currently an upper limit, 
barely providing the constraint,
we assume $F^{D^+\to S_{ud}}=\sqrt 2 F^{D^+\to S_{ns}}$ 
under $SU(3)_f$ symmetry for a practical extraction. Therefore,
observing $D^+\to f_0 e\nu,f_0\to\pi^+\pi^-$ is crucial for
constraining $F^{D^+\to S_{ns}}$ and $F^{D^+\to S_{ud}}$, 
and testing possible $SU(3)_f$ symmetry breaking.

As we establish a solid statistical and theoretical foundation
for discerning the nature of the light scalar meson,  it might appear that
the molecule scenario is ignored in this study.
In fact, the molecule scenario cannot be as relevant in weak decays as it is
in meson-meson interactions, thereby 
giving a suppressed contribution to $D\to S_0 e\nu, S_0\to M_1 M_2$.

We use $D^+_s\to \hat f_0 e\nu,\hat f_0\to \pi^+ \pi^-$ to explicitly estimate the suppression, 
with $\hat f_0$ denoted as a $K\bar K$ molecular component of $f_0(980)$~\cite{Branz:2007xp}.
The molecular state typically requires the residual nuclear force to provide
additional binding after the meson pair is produced.
However, since QCD and gluon exchanges are responsible for the hadronization of a final state
in weak decays, rather than the residual nuclear force, 
the resonant $D^+_s\to \hat f_0 e^+\nu_e,\hat f_0\to \pi^+ \pi^-$ decay does not allow 
the {\it ``direct''} $D^+_s\to \hat f_0$ transition~\cite{Hsiao:2019ait,Yu:2021euw,Wang:2021ews}.
One possible way to form $\hat f_0$ involves the non-resonant $D^+_s\to K\bar K$ transition 
with $K\bar K=K^+ K^-$ or $K^0\bar K^0$, followed by the $K\bar K$-loop to $\hat f_0$ 
as the final state interaction~\cite{Hsiao:2019ait,Yu:2021euw,Wang:2021ews,Sekihara:2015iha},
and then $\hat f_0$ decays into $\pi^+\pi^-$ after resonating for a while.

Clearly, the molecule state leads to a more complex decay process.
To estimate the molecular contribution, we first evaluate the non-resonant branching fraction 
${\cal B}_{\rm NR}(D_s^+\to K \bar K e^+\nu_e)\simeq 1\times 10^{-3}$,
which is based on 
${\cal B}_{\rm NR}(D_s^+\to K^+ K^- e^+\nu_e)
\simeq (\tau_{D_s^+}/\tau_{D^+}){\cal B}_{\rm NR}(D^+\to \pi^+ K^- \mu^+\nu_\mu)$
with $SU(3)_f$ symmetry and the observed
${\cal B}_{\rm NR}(D^+\to \pi^+ K^- \mu^+\nu_e)=(1.9\pm 0.5)\times 10^{-3}$~\cite{pdg}.
We thus obtain
\begin{eqnarray}\label{est}
{\cal B}(D_s^+\to \hat f_0 e^+\nu_e,\hat f_0\to \pi^+\pi^-)<2.5\times 10^{-4}\,,
\end{eqnarray}
by using ${\cal B}_{\rm NR}(D_s^+\to K\bar K e^+\nu_e)$ 
and accounting for the $\hat f_0\to\pi^+\pi^-$ resonant effect,
with $|\hat f_0\rangle=(|K^+ K^-\rangle+|K^0\bar K^0\rangle)/\sqrt 2$~\cite{Branz:2007xp}. 
Meanwhile, the $K\bar K$-loop to $\hat f_0$ calculation introduces another suppression factor, 
resulting in the less-than notation in Eq.~(\ref{est}). 
It is noteworthy that our estimate aligns, to the order of magnitude, 
with the value of $5.1\times 10^{-4}$ 
obtained from calculating the molecular contribution~\cite{Sekihara:2015iha}.
Since the molecular state is expected to make a suppressed contribution,
the semileptonic $D$ decay serves as a specific investigation 
of the $Z$-included $q\bar q$ and $q^2\bar q^2$ structures.

The $SU(3)_f$ symmetry was previously applied to $B\to J/\Psi S_0$~\cite{Stone:2013eaa}, 
semileptonic $D$ decays~\cite{Wang:2009azc}, and non-leptonic $D$ decays~\cite{Cheng:2002ai}
for exploring the scalar mesons. Accordingly, 
LHCb presented~\cite{LHCb:2014ooi,LHCb:2014vbo}
\begin{eqnarray}
{\cal R}(\bar B^0_s)\equiv
\frac{{\cal B}(\bar B^0_s\to J/\psi \sigma_0,\sigma_0\to\pi^+\pi^-)}{{\cal B}(\bar B^0_s\to J/\psi f_0,f_0\to\pi^+\pi^-)}
<3.4\%\,,\nonumber\\
{\cal R}(\bar B^0)\equiv\frac{{\cal B}(\bar B^0\to J/\psi f_0,f_0\to\pi^+\pi^-)}{{\cal B}(\bar B^0\to J/\psi \sigma_0,\sigma_0\to\pi^+\pi^-)}
<1.8\%\,.
\end{eqnarray}
While ${\cal R}(\bar B^0_s)$ is translated into $|\theta|<7.7^\circ$, consistent with the tetraquark picture,
${\cal R}(\bar B^0_s)$ leads to a larger limit of $|\theta|<17^\circ$~\cite{LHCb:2014vbo}, 
claimed to be inconsistent with the tetraquark picture. The inconsistency might indicate that
the detailed information about 
the $\bar B^0_{(s)}$ to $(f_0,\sigma_0)$ transition form factors should be considered,
such as the $SU(3)_f$ relations, the momentum dependence, the mixing scenarios 
in the $q\bar q$ and $q^2\bar q^2$ structures, and the resonant effects, 
as we have performed in semileptonic $D_s^+$ and $D^{+,0}$ decays.

In summary, 
we have developed a highly sensitive analysis for distinguishing 
between the quark-antiquark and tetraquark nature of the scalar mesons
in semileptonic decays of $(D^{+,0}, D^+_s)\to S_0 e^+\nu_e, S_0\to M_1 M_2$.
A comprehensive global fit was conducted to accommodate existing data,
subsequently extracting the $D\to a_0, f_0, \sigma_0$ transition form factors 
in the two different quark contents for the first time. Specifically, our calculations yielded 
${\cal B}(D_s^+\to \sigma_0 e^+\nu_e,\sigma_0\to\pi^+\pi^-)=(20.3\pm 1.8\pm 0.5)\times 10^{-4}$ 
for the $q\bar q$ structure and $(0.58^{+1.43}_{-0.57}\pm 0.01)\times 10^{-4}$ for the $q^2\bar q^2$ structure.
The branching fraction in the $q\bar q$ structure
exhibited significant $9\sigma$ deviations from the experimental upper limit of $3.3\times 10^{-4}$,
indicating that even exhaustive consideration of all possible uncertainties cannot accommodate the deviations.
In contrast,
the branching fraction in the $q^2\bar q^2$ structure fell within the experimental allowed range.
Therefore, we have clearly tested $S_0$ as a non-$q\bar q$ state.

\section*{ACKNOWLEDGMENTS}
YKH was supported in parts by NSFC (Grants No.~12175128 and No.~11675030).
BCK was supported in parts  by NSFC (Grants No.~11875054 and No.~12192263),
Joint Large-Scale Scientific Facility Fund of the NSFC and CAS (Grant No. U2032104), and 
the Excellent Youth Foundation of Henan Scientific Committee under Contract No.~242300421044.

\end{document}